\documentclass[conference, twocolumn]{IEEEtran}
\usepackage{amsmath,amssymb,graphicx,epsfig,fancyhdr,amsthm,tabulary,epstopdf}
\usepackage{hyperref}
\usepackage{cite}
\usepackage{multirow}
\usepackage{float}
\usepackage{amsfonts}
\usepackage{footnote}
\makesavenoteenv{tabular}
\usepackage{tabulary,bigstrut,array,multirow,booktabs}%
\usepackage{tikz}
\usetikzlibrary{calc}
\usepackage{pdfsync}
\usepackage{bm}
\theoremstyle{remark}
\newtheorem{rem}{Remark}

\theoremstyle{definition}

\newcommand{\rank}{\mathop{\mathrm{rank}}}

\begin{document}
%\pagestyle{fancy}
%\fancyhead[LO,LE]{\sf McGill University\\Electrical and Computer Engineering}

%%%%%%%\title{On the Capacity of the Cognitive Gaussian Z-Interference Channel:\\ Less Noisy and More Capable CGZIC}
\title{Extended Subspace Error Localization for Rate-Adaptive Distributed Source Coding}
%\\ Extended Subspace Error Localization for Rate-Adaptive distributed source coding based on DFT Codes
%\\ Systematic DFT Frames: Applications and Extreme Eigenvalues (Properties)}

\author{\IEEEauthorblockN{Mojtaba Vaezi and Fabrice Labeau}\\
\IEEEauthorblockA{Department of Electrical and Computer Engineering\\
McGill University,
Montreal, Quebec H3A 0E9, Canada\\
Email: mojtaba.vaezi@mail.mcgill.ca, fabrice.labeau@mcgill.ca}
%\and
%\IEEEauthorblockN{Mai Vu}
%\IEEEauthorblockA{Department of Electrical and Computer Engineering\\
%McGill University\\
%Montreal, Quebec H3A 2A7, Canada\\
%Email: mai.h.vu@mcgill.ca}
}


\maketitle

%----------------------------------------------------------------------
\begin{abstract}
A subspace-based approach for rate-adaptive distributed source coding (DSC)
based on discrete Fourier transform (DFT) codes is
developed.
Punctured DFT codes can be used to implement rate-adaptive
source coding, however they perform poorly after even moderate puncturing since
the performance of the subspace error localization degrades severely.
%Further, puncturing can only be used for the parity-based DSC.
The proposed subspace-based error localization
 extends and improves the existing one, based on
additional syndrome, and is naturally suitable for rate-adaptive distributed
source coding architecture.
%It is applied both to the syndrome- and parity-based DSC.

\end{abstract}

%\begin{keywords}
%Tight frames,  BCH-DFT codes, eigenvalue.
%\end{keywords}

\IEEEpeerreviewmaketitle

\section{Introduction}
{\let\thefootnote\relax\footnotetext{This work was supported by Hydro-Qu\'{e}bec,
the Natural Sciences and Engineering Research Council of Canada and McGill
University in the framework of the NSERC/Hydro-Qu\'{e}bec/McGill Industrial
Research Chair in Interactive Information Infrastructure for the Power Grid.}}

The ideas of {\it coding theory} can
be described within the setting of {\it signal processing}
by using a class of real (or complex) Bose-Chaudhuri-Hocquenghem (BCH) codes
\cite{blahut2003algebraic} known as the discrete Fourier transform codes.
 {\it DFT codes} find applications in different areas including
wireless communications \cite{wang2003complex}, joint source-channel coding
\cite{gabay2007joint}, and distributed source coding \cite{Vaezi2011DSC}.
Looking from a {\it frame theory} perspective, these codes are used to provide robustness to {\it erasure}
in wireless networks  \cite{goyal2001quantized,rath2004frame,bodmann2011burst}.

When {\it error} correction is required \cite{wang2003complex,gabay2007joint,Vaezi2011DSC},
{\it error localization} is a crucial step of the decoding algorithm of DFT codes.
Error localization in BCH-DFT codes can be done by extending
that of binary BCH codes to the real field \cite{blahut2003algebraic}.
Rath and Guillemot \cite{rath2004subspace} used {\it subspace-based}
error localization and showed that it outperforms the {\it coding-theoretic}
approach; the improvement is achieved by mitigating the effect of  the quantization error
by involving as many syndrome samples as possible.
The authors recently employed  DFT codes for lossy DSC \cite{Vaezi2011DSC}
and adopted subspace error localization in this context \cite{Vaezi2012WZ}.
This approach to DSC exploits the correlation between the sources in the {\it analog} domain and
it is promising in delay-sensitive applications.
%, as using short codes it can outperform DSC based on binary codes for specific channels, e.g., when correlation between the sources is an impulsive channel.
The performance of the system, like other DSC systems,
degrades when the correlation between the sources is unstable.
Although {\it puncturing} can be used for {\it rate-adaption},
it severely affects the error localization
 and substantially increases the end-to-end distortion.

The primary contribution of this paper is to
develop {\it rate-adaptive}
distributed source codes based on DFT codes.
%that have better performance than alternative codes of linear complexity encoding and decoding.
To do so, we extend and improve the subspace
error localization of DFT codes and adapt it both to the parity- and syndrome-based DSC.
The extended subspace error localization is  applicable
to other codes based on orthogonal transform matrices such as the
discrete cosine transform (DCT) and discrete sine transform (DST) codes, as the subspace approach is \cite{kumar2010improved}.

Rate-adaptation for the parity approach is
developed only for real DFT codes of rate $0.5$. However, the syndrome approach is
for any real or complex code; the encoder transmits
a short syndrome based on an $(n,k)$ code and augments
it with additional samples if decoding failure is fed back.
The algorithm is incremental so that there is no need to re-encode
the sources when more syndrome is requested.

The paper is organized as follows. After a brief review of DFT codes in Section~\ref{sec:DFTcodes},
we discuss how the subspace error localization outperforms the coding-theoretic approach
and introduce the extended subspace decoding in Section~\ref{sec:Loc}.
We explain the rate-adaptive DSC system in Section~\ref{sec:DSC}.
Numerical results in Section~\ref{sec:sim} confirm the merit of the proposed error localization.
This is followed by conclusion in Section~\ref{sec:con}.

\section{DFT Codes}
\label{sec:DFTcodes}

The generator matrix of an $(n,k)$ DFT code \cite{Marshall}, in general, consists of
any $k$ columns of the inverse DFT (IDFT) matrix of order $n$;
the remaining $n-k$ columns of this matrix are used to build the {\it parity-check matrix} $H$.
These codes are a family of {\it cyclic codes} over the complex field.
Thus, their codewords satisfy certain spectral properties in the frequency domain \cite{blahut1992algebraic}.
Within the class of DFT codes, there are BCH codes in the complex and real fields.
%More specifically,
Each codeword of an $(n,k)$ BCH-DFT code
has $d\triangleq n-k$ cyclically adjacent zeros in the frequency domain. They are
{\it maximum distance separable} codes with {\it minimum Hamming distance} $d_{\min}=d+1$.
They are, hence, capable of correcting up to $t=\lfloor \frac{d}{2}\rfloor$ errors.

We consider real BCH-DFT codes whose generator matrix, for an $(n,k)$ code, is defined by  \cite{vaezi2012frame,gabay2007joint}
\begin{align}
G= \sqrt{\frac{n}{k}} W_n^H \Sigma W_k,
\label{eq:G1}
\end{align}

\noindent where $W_n$ and $W_k$ are the DFT matrices of size $n$ and $k$, and
\begin{align}
\Sigma= \left( \begin{array}{ccccccc}
       I_\alpha  & \bm{0}  \\
       \bm{0}   & \bm{0}  \\
      \bm{0}    &   I_\beta    \\
      \end{array}
\right) \label{eq:cov}
\end{align}
is an $n\times k$ matrix with $\alpha = \lceil \frac{n}{2}\rceil  -\lfloor \frac{n-k}{2}\rfloor$
and $\alpha + \beta =k$.

%
%In general, every sample in the codewords of a DFT code is a linear
%combination of all data samples of the input block, i.e., the data samples do not appear explicitly in the codewords.
%A specific method of encoding, known as {\it systematic encoding}, leaves the data samples unchanged.
%These unchanged samples can be exhibited in any $k$ components of the codeword, but
% evenly-spaced data samples minimizes the variance of parity
%samples, for a given input variance \cite{vaezi2012frame}. Such a code reduces the required transmission rate
%for a fixed distortion \cite{Cover, vaezi2012frame}.
%Systematic codes are particularly important in the context of distributed source coding \cite{Vaezi2011DSC}.

The generator matrix of a complex BCH-DFT code can be achieved by removing $W_k$ from \eqref{eq:G1};
we can also remove the constraint on $\alpha$.
Although we focus on the real BCH-DFT codes, the results we present
in this paper are valid for the complex codes as well.
In the rest of this paper, for brevity, BCH-DFT codes will be referred to as DFT codes.

\section{Error Localization in DFT codes}
\label{sec:Loc}

Let $\bm{r} = \bm{c} + \bm{e}$ be a noisy version of codeword $\bm{c}$
 generated by a DFT code and suppose that the error vector $\bm{e}$ has $\nu \leq t$
 nonzero elements. Let $ i_1, \hdots, i_{\nu}$ and $e_{i_1}, \hdots, e_{i_{\nu}}$, respectively,
denote the locations and magnitudes of the nonzero elements.
The decoding algorithm in DFT codes is composed of
three main steps~\cite{blahut2003algebraic}: {\it error detection} (to determine  $\nu$),
{\it error localization} (to find $ i_1, \hdots, i_{\nu}$), and
{\it error calculation} (to calculate $e_{i_1}, \hdots, e_{i_{\nu}}$).
This section is focused on the error localization. Thus,
we assume that the number of errors $\nu$ is known at the decoder.

The syndrome of $\bm{e}$, a key for the decoding algorithm, is computed as
\begin{align}
 \bm{s}= H\bm{r}= H(\bm{c} + \bm{e})= H\bm{e},
\label{eq:synd}
\end{align}
where $\bm{s}=[s_1,\, s_2, \hdots, s_d]^T$  is a complex vector with
\begin{align}
s_m=\frac{1}{\sqrt{n}}\sum_{p=1}^{\nu} e_{i_p}X_p^{\alpha-1+m}, \quad m=1,\hdots, d,
\label{eq:syndsamp}
\end{align}
in which $\alpha$ is defined in \eqref{eq:cov} and $X_p= e^{ \frac{j2\pi i_p}{n}}$, $ p=1, \hdots, \nu.$

\subsection{Coding-Theoretic and Subspace Approaches}\label{sec:SS}
The classical approach to the error localization is to
identify an  {\it error locator polynomial} whose roots correspond
to error locations.
 The  error locator polynomial is defined as
\begin{align}
\Lambda(x) = \prod_{\substack{i=1}}^{\nu} (1-xX_i) = 1+ \Lambda_1 x + \cdots+ \Lambda_\nu x^\nu,
\label{eq:poly}
\end{align}
 and its roots $X_1^{-1}, \hdots, X_\nu^{-1}$ correspond to the error locations
 $i_p, p \in [1,\hdots,\nu]$, as $X_{p}^{-1} = \omega^{i_p}$ where  $\omega= e^{-j\frac{2\pi}{n}}$.
The coefficients $\Lambda_1, \hdots, \Lambda_\nu$  can be found by solving the following set
 of {\it consistent} equations \cite{blahut2003algebraic}
\begin{align}
s_j \Lambda_{\nu} + s_{j+1} \Lambda_{\nu-1} +  \cdots + s_{j+\nu -1} \Lambda_{1}= - s_{j+\nu},
\label{eq:synd2}
\end{align}
 for  $j=1, \hdots, d-\nu$.
To put it differently, as the IDFT of $\bm{\Lambda}_n = [1, \Lambda_1,\hdots,\Lambda_\nu,\bm{0}_{1 \times (n-\nu -1)} ]^T$
becomes zero at the error locations, the circular convolution of $\bm{\Lambda}_n$ with the DFT of the 	
error vector is a zero vector \cite{blahut2003algebraic, rath2004subspace}.

An alternative approach is to use the {\it subspace} methods for error localization  \cite{rath2004subspace}.
The {\it error-locator matrix} of order $m$,
whose columns are the {\it error-locator vectors} of order $m$, is a {\it Vandermonde} matrix defined as
\begin{align}	
V_m  = \left[ \begin{array}{cccc}
 1 & 1 & \hdots & 1  \\
 X_1 & X_2 & \hdots & X_{\nu}  \\
 \vdots & \vdots & \ddots & \vdots\\
X_{1}^{m-1} & X_{2}^{m-1} & \hdots & X_{\nu}^{m-1}  \\  \end{array}
\right].
\label{eq:Vander}
\end{align}
Next, following the nomenclature of \cite{rath2004subspace}, for $\nu+1 \le m \le d-\nu+1$, we define the syndrome matrix by
\begin{align}
S_m = V_m D V_{d-m+1}^{T},
\label{eq:S}
 \end{align}
where $D$ is a diagonal matrix of size $\nu$ with nonzero diagonal elements
 $d_{p} = \frac{1}{\sqrt{n}} e_{i_p}X_{p}^{\alpha},  p=1, \hdots, \nu.$
One can check that
 \begin{align}S_m&=  \left[ \begin{array}{cccc}
 s_1 & s_2 & \hdots &s_{d-m+1}  \\
 s_2 & s_3 & \hdots & s_{d-m+2}  \\
 \vdots & \vdots & \ddots & \vdots\\
s_{m} & s_{m+1} & \hdots & s_{d}  \\  \end{array}
\right].
\label{eq:syndmatrix}
 \end{align}

\noindent Also, we define the covariance matrix as
\begin{align}
R_m=S_mS_m^H.
\label{eq:R}
\end{align}
From \eqref{eq:S}, it is obvious that the rank of $R_m $ is $\nu$; thus, it can be eigendecomposed as
\begin{align}
R_m = [ U_e \; U_n]\left[ \begin{array}{cc}
 \Delta_e & \bm{0}   \\
\bm{0} &  \Delta_n \\
 \end{array}
\right] [ U_e \; U_n]^{H},
\label{eq:eigen}
 \end{align}
where the square matrices $\Delta_e$ and $\Delta_n$ contain the $\nu$ largest
and $m- \nu$ smallest eigenvalues, and $U_e$ and $U_n$
contain the eigenvectors corresponding to $\Delta_e$ and $\Delta_n$,
respectively.\footnote{Clearly, since there is no noise (or quantization error),
$\Delta_n = \bm{0}$  and $\Delta_e$ contains the $\nu$ nonzero eigenvalues of $R_m$.}
The sizes of  $U_e$ and $U_n$ are $m \times \nu$ and $m \times (m-\nu)$.
The columns in $U_e$ span the {\it channel-error subspace} spanned by $V_m$ \cite{rath2004subspace}.
Thus, the columns in $U_n$ span the {\it noise subspace}.
Then, from the fact that $U_e^H U_n = \bm{0}$, we conclude that
\begin{align}
V_m^H U_n = \bm{0}.
\label{eq:VU}
 \end{align}
Now, let $\bm{v} = [1, x, x^2,\hdots,x^{m-1}]^T$ where $x$ is a complex variable %and $\bm{u}_{n,j}$ be the $j$th column of $U_n$
and define the function
\begin{align}
%F(x) \triangleq \bm{v}^H U_nU_n^H \bm{v} =  \sum_{j=1}^{m-\nu} \sum_{i=0}^{m-1} f_{ji}|x^2|^{i}.\\
F(x) \triangleq \sum_{j=1}^{m-\nu} \bm{v}^H U_n =  \sum_{j=1}^{m-\nu} \sum_{i=0}^{m-1} f_{ji}x^{i}.
%F(x) \triangleq \sqrt{\bm{v}^H U_nU_n^H \bm{v}} =  \sum_{j=1}^{m-\nu} \sum_{i=0}^{m-1} f_{ji}x^{i}. \\
%F(x) \triangleq \sqrt{\bm{v}^H U_nU_n^H \bm{v}}  =  \sum_{j=1}^{m-\nu} \sum_{i=0}^{m-1} f_{ji}|x|^i.\\
%F(x) \triangleq \bm{v}^H U_nU_n^H \bm{v} =  \sum_{j=1}^{m-\nu} \sum_{i=0}^{m-1} |f_{ji}x^i|^2.
\label{eq:F}
 \end{align}
$F(x)$ can be considered as sum of $m-\nu$ polynomials $\{f_j\}_{j=1}^{m-\nu}$ of order $m-1$; each polynomial
corresponds to one column of $U_n$. Let $\mathcal{F}$ denote this set of polynomials.
In light of \eqref{eq:VU}, each one of these polynomials vanishes for
$x= X_1, \hdots, X_\nu$, i.e., $F(x)=0$  for $X_1, \hdots, X_\nu$.
These are the only common roots of $\{f_j\}$
over the $n$th roots of unity  \cite{rath2004subspace}; thus, the errors location can be determined by finding
the zeros of $F(x)$ over the set of $n$th roots of unity.  Equivalently,
one may use the {\it signal subspace} to find the error location \cite{kay1988modern}.
%, as in general either signal subspace or noise subspace  can be used for estimation

The subspace method outperforms the coding theoretic error localization.
To prove this, we can see that $\Lambda(x)$ is the smallest degree polynomial that has roots in
 $X_1, \hdots, X_\nu$ and lies in the noise subspace; it is achieved for $m=\nu +1$
 in \eqref{eq:F}. As $m$ increases the degree of polynomials $\{f_j\}$ goes up which
 gives more degrees of freedom and helps improve the estimation of roots, and the error locations consequently.
 Another factor that affects location estimation is the number of
 polynomials $\{f_j\}$ with linearly independent coefficients.
 The more there are such polynomials, the better the estimation is as the variations due to noise (quantization)
 are reduced by adding such independent polynomials in $F(x)$.

 Although the number of
 polynomials increases with $m$, their coefficients may not be independent.
 The latter depends on the number of nonzero eigenvalues in the noise subspace which is,
 in turn, related to the rank of $S_m$ and is limited by
 \begin{equation}
\begin{aligned}
\rank(S_m) \le \underset{m}{\operatorname{max}}\operatorname{min} (m, d-m+1) = \left \lceil \frac{d}{2} \right \rceil.
\end{aligned}
\label{eq:Opt}
\end{equation}
This suggests that the optimum value for $m$ is $\lceil \frac{d}{2}\rceil$.
Then, from \eqref{eq:F}, one can check that the subspace approach will
result in a better  error localization than the coding-theoretic approach,
except when $\nu = t$ and $d$ is even; in this latter case $m = \nu +1$
and there is just one polynomial and its degree is $ \nu $, the same
as \eqref{eq:poly} in the coding-theoretic approach.

In practice, where quantization comes into play, the received vector
 is distorted both by the error vector $\bm{e}$ and
quantization noise $\bm{q}$. Therefore $\bm{r} = \bm{ c} + \bm{e} +\bm{q}$, and
 its syndrome is only a perturbed version of $\bm{s}$  because
\begin{align}
H\bm{r}= H(\bm{ c} + \bm{q}+ \bm{e})= \bm{s}_q + \bm{s} = \tilde{\bm{s}},
\label{eq:syndq}
\end{align}
where $\bm{s}_q \equiv H\bm{q}$ and  $\bm{q}=[q_1,\, q_2, \hdots, q_n]^T$ is the quantization error.
The distorted syndrome samples can be written as
\begin{align}
\tilde{s}_m = \frac{1}{\sqrt{n}}\sum_{p=1}^{\nu} e_{i_p}X_p^{\alpha -1 + m} +  \frac{1}{\sqrt{n}}\sum_{p'=1}^{n} q_{i_{p'}}X_{p'}^{\alpha -1 + m},
\label{eq:syndq2}
\end{align}
where $i_{p'}$ shows the index for quantization error. The distorted syndrome matrix $\tilde{S}_m$  and its
corresponding covariance matrix $\tilde{R}=\tilde{S}_m\tilde{S}_m^H$
are defined similar to \eqref{eq:syndmatrix} and \eqref{eq:R} but for the distorted syndrome samples.

\subsection{Extended Subspace Approach}\label{sec:ExSS}

The main idea behind the extended subspace approach is to enlarge the
dimension of the noise subspace such that, in \eqref{eq:F}, the  number of
 polynomials with linearly independent coefficients
 and/or their degree grow.
 This can be accomplished by constructing an extended syndrome matrix $S'_m $,
 in the form of $S_m $ in \eqref{eq:syndmatrix} but for $  d' > d$, which is decomposable as
\begin{align}
S'_m = V_m D V_{d'-m+1}^{T},
 \label{eq:S'}
 \end{align}
for $\nu+1 \le m \le d'-\nu+1$, and  $V_m$ and $D$ defined in \eqref{eq:S}.
Following the same argument that led to \eqref{eq:Opt}, it is easy to see that
the optimal $m$ is $\lceil \frac{d'}{2}\rceil$. Then, as explained in Section~\ref{sec:SS},
this will improve the error localization.

 To form $S'_m$, we first define the extended syndrome $\bar{\bm{s}}$.
 Let $ \bar d  \in [d+1, n]$ show the new number of syndrome samples where
 there are $\bar d-d$ additional samples as compared to \eqref{eq:syndsamp}.
 Similar to the syndrome vector $\bm{s}$, we define the extended
 syndrome vector $\bar{\bm{s}}$  as
 \begin{align}
\bar{\bm{s}}= \bar{H}\bm{r}=  \bar{H}\bm{c} +\bar{H}\bm{e},
\label{eq:Esynd}
\end{align}
where $\bar{H}$ consists of those
$k$ columns of the IDFT matrix of order $n$ used to build $G$.
%In other words, $\bar{H}$ is the complement of $H$.
%Therefore, $\bar{\bm{s}}=[\bar{s}_1,\, \bar{s}_2, \hdots, \bar{s}_{n-d}]^T$ and
More precisely, for $m=1,\hdots, \bar d-d$,
\begin{align}
\bar{s}_m=\frac{1}{\sqrt{n}}\sum_{p=1}^{\nu} e_{i_p}X_p^{d+\alpha-1+m} +  \frac{1}{\sqrt{n}}\sum_{p'=1}^{n} c_{i_{p'}}X_{p'}^{d+\alpha -1 + m}.
\label{eq:syndsamp1}
\end{align}
 Now, with
\begin{align}
s'_m=  \left\{
  \begin{array}{l l}
    s_m, & \quad  1\le m\le d,\\
     \bar{s}_{m-d}, &  \quad d< m\le d',
     \label{eq:consts}
  \end{array} \right.
  \end{align}
$S'_m $ will be decomposable as  \eqref{eq:S'} provided that
the second term in the right-hand side of  \eqref{eq:syndsamp1} is vanished, or equivalently  $ \bar{H}\bm{c}$ is removed from \eqref{eq:Esynd}.
Observe that considering  quantization $c_{i_{p'}}$ will be  replaced by  $c_{i_{p'}} +q_{i_{p'}}$; i.e.,  $\tilde{\bar s}_m$
contains a term related to quantization error, similar to $\tilde{s}_m$ in  \eqref{eq:syndq2}.
Likewise, $\tilde{s}'_m$ is built upon $\tilde{s}_m$ and $\tilde{\bar s}_m$.
Again we should emphasize that using
$R'_m=S'_m S'^H_m$ (and $\tilde R'_m$) may not necessarily improve the error localization;
to expect  gain by virtue of the extended subspace method,
 we need to compensate for the term $\bar{H}\bm{c}$ in \eqref{eq:Esynd}.
 This is done for the syndrome-based DSC
in the next section.

\begin{figure}
  \centering
 \includegraphics [scale=0.46] {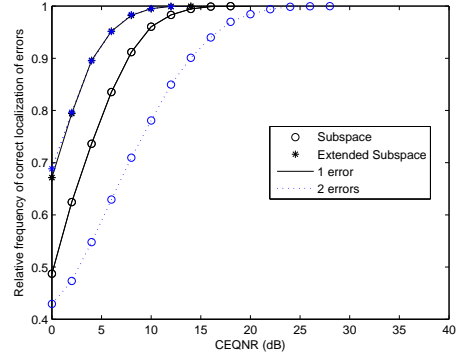}
  %\centering{\epsfig{figure=IC.eps, width=8cm}}
  \caption{Probability of correct localization of
   errors using the subspace and extended subspace approaches, at different channel-error-to-quantization-noise ratio (CEQNR), for a $(10, 5)$ DFT code
  where all errors are in even positions.}
  \label{fig:fig1}
    \vspace{-.1cm}
\end{figure}

Before moving on to the next section, we look at extended subspace method for a special, yet important,
class of DFT codes where $n=2k$. For such a code, $d=k$ and $X_p^d$ is
$+1$ ($-1$) for errors in the even (odd) positions in the codeword. Then,
if all errors are in the even (odd) positions\footnote{Although this condition might seem unrealistic at first glance,
in the next section we show that it is realized, for instance, in a parity-based DSC.}, we can simply replace $\bar{\bm{s}}$ with
$\bm{s}$ ($-\bm{s}$). Thus, using \eqref{eq:consts} we can form $S'_m (\tilde S'_m)$ and the corresponding $R'_m (\tilde R'_m)$.
Subsequently, the eigendecomposition of $\tilde R'_m$ for $m=\lceil d'/2 \rceil$
increases the number of polynomials in $\mathcal{F}$ and their degree.
Figure~\ref{fig:fig1}  shows the merit of extended error localization to the existing one, for $d'=n$, in  a $(10,5)$ code.
Such a big gain in error localization is achieved by using the same $d$ syndrome samples but
forming a larger syndrome matrix  which allows a larger noise subspace.

\begin{rem}
Similar to the subspace approach \cite{kumar2010improved}, the extended subspace approach can be applied to the DCT and DST codes; further, it can be used  even for  the
non-BCH DCT and DST codes \cite{kumar2010improved}.
\end{rem}

\begin{rem}
Knowing that $\tilde R_m$ can also be used to determine the number of errors $\nu$ \cite{Vaezi2012WZ},
where the extended error localization is applicable, $\tilde R'_m$ can be used for this purpose and it improves the
results reasonably.
\end{rem}

\section{Rate-Adaptive Distributed Lossy Source Coding Using DFT Codes}
\label{sec:DSC}

Distributed {\it lossless} compression of two correlated sources can be as efficient
as their joint compression \cite{SW}. This is also valid for {\it lossy} source
coding  with side information at the decoder
 for jointly Gaussian sources and the mean-squared error (MSE) distortion
measure \cite{WZ}. Tipically, DSC is realized by quantizing the sources and
applying Slepian-Wolf coding in the binary domain.
Slepian-Wolf coding can be implemented in the analog domain as well \cite{Vaezi2011DSC} which
outperforms its binary counterpart for certain scenarios, e.g., an impulsive correlation model.
The proposed DSC schemes based on  DFT codes, both parity  and
syndrome approaches, are also appropriate
for low-delay coding as they perform sufficiently well even when short
source blocks are encoded.

When the statistical dependency between the sources varies or is not known at the encoder,
a {\it rate-adaptive} system with feedback is an appealing
solution \cite{varodayan2006rate}.
Rate-adaptive DSC based on binary codes, e.g., puncturing the parity or syndrome bits of
turbo and LDPC codes, have been proposed in \cite{varodayan2006rate,toto2008rate}.
In the sequel, we extend DSC  based on DFT codes \cite{Vaezi2011DSC} to perform DSC
in a rate-adaptive fashion.
We consider two continuous-valued correlated sources $\bm x$ and $\bm y$
where $x_i$ and $y_i$ are statistically dependent  by
 $y_i=x_i + e_i$, and $e_i$ is continuous, i.i.d.,
and independent of $x_i$.
%We are interested in the compression of $\bm x$
%given that $\bm y$ is known at the decoder, only.

\subsection{Parity-Based Approach }\label{sec:par}
Puncturing is a well-known technique used to
achieve higher rate codes for the same decoder;
%it increases the flexibility
%of the system without significantly increasing its complexity.
%To puncture a code, we delete columns from its generator matrix.
it is inherently well-suited for parity-based DSC schemes as
one can remove some of the parity samples to puncture a code.
However, with  subspace
error localizations, the performance of punctured DFT codes deteriorates largely.
As explained in the
previous section, extended subspace decoding significantly improves
the results provided that the errors are restricted to even (odd) positions.
This can be achieved by using $(2k,k)$ DFT codes. Generated by \eqref{eq:G1}, a $(2k,k)$ DFT code
is systematic with parity samples in even positions. We can modify this code and
form a code whose parity samples are in
the even positions \cite{vaezi2012frame}. Then, the extended syndrome matrix in  \eqref{eq:S'}
can be used both for error detection and localization.
Similar to the subspace method, the performance of the system drops sharply with puncturing.
Furthermore, although simple, puncturing may cause the minimum distance to decrease.
%It is also restricted to rate $1/2$ codes if we strive to apply the
%extended error localization.
An alternative, general approach for rate-adaptation is presented next.

\subsection{Syndrome-Based Approach }\label{sec:synd}

 Rate-adaption using puncturing is not natural for syndrome-based DSC
systems \cite{toto2008rate}. Instead, the encoder can transmit a short
syndrome based on an aggressive code and augment it
with additional syndrome samples,  if decoding fails.
This process loops until the decoder gets sufficient syndrome
for successful decoding. This approach
is viable only for feedback channels with reasonably short round-trip time \cite{varodayan2006rate}.

In the syndrome-based DSC based on DFT codes \cite{Vaezi2011DSC},
the encoder computes $\bm{s}_x$ and transmits it to the decoder.
At the decoder, we have access to the side information $\bm{y}=\bm{x}+\bm{e}$ and can compute
its syndrome so as to find $\bm{s}_{e}=\bm{s}_{y}- \bm{{s}_{x}}$. For rate adaptation, if needed, the encoder
transmits $\bar{\bm{s}}_x= \bar{H}\bm{x}$ sample by sample; the receiver also can compute
$\bar{\bm{s}}_y= \bar{H}\bm{y}= \bar{\bm{s}}_x + \bar{\bm{s}}_e$ and evaluate
$ \bar{\bm{s}}_e = \bar{\bm{s}}_y - \bar{\bm{s}}_x$. After that, we can form the
extended syndrome matrix $S'_m $  by
replacing  $\bm{s} = \bm{s}_e $ and $ \bar{\bm{s}} = \bar{\bm{s}}_e $ in the right-hand side of \eqref{eq:consts}.
Clearly, when quantization is considered this equation needs to be updated as
\begin{align}
\tilde s'_m=  \left\{
  \begin{array}{l l}
    \tilde s_m, & \quad  1\le m\le d,\\
     \tilde{\bar{s}}_{m-d}, &  \quad d< m\le n,
     \label{eq:consts2}
  \end{array} \right.
  \end{align}
in which $\tilde {\bm{s}} = \bm{s}_e + \bm{s}_q $, $ \tilde{\bar{\bm{s}}} = \bar{\bm{s}}_e + \bar{\bm{s}}_q$,
and $\bar{\bm{s}}_q= \bar{H}\bm{q}$.

The new $\tilde R'_m=\tilde S'_m \tilde S'^H_m$ then is used for error localization as detailed in Section~\ref{sec:Loc}.
Note that the code is incremental, so the encoder does not need to re-encode the sources when more
syndrome is requested.
It buffers and transmits syndrome to the decoder sample by sample.
Moreover, we can use $\tilde R'_m$
to find the number of errors as explained  in  \cite{Vaezi2012WZ}.

\section{Simulation Results}
\label{sec:sim}

To evaluate the performance of the algorithm
we do simulation using
a Gauss-Markov source with mean zero, variance one, and
correlation coefficient 0.9 for two DFT codes, namely, $(10, 5)$ and $(17, 9)$.
For each code, we generate the syndrome and
extended syndrome, quantize them with a 3-bit uniform quantizer with step size $\Delta= 0.25$, and
transmit them over a noiseless communication media.
We plot the relative frequency of correct localization of different numbers of errors.
To do so, we define channel-error-to-quantization-noise ratio (CEQNR)
as the ratio of channel error power to the quantization noise power ($\sigma_e^2/\sigma_q^2$) and,
similar to \cite{rath2004subspace}, we assume that the channel error components are fixed.
The number of errors in each block is limited to $t$, the error correction capacity of the code.
The simulation results are for $10^4$ input blocks for each CEQNR.

\begin{figure}
  \centering
 \includegraphics [scale=0.488] {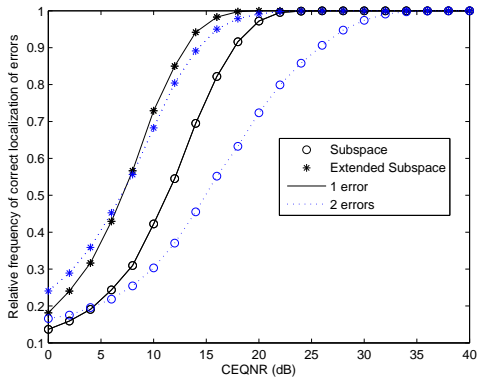}
  %\centering{\epsfig{figure=IC.eps, width=8cm}}
  \caption{Probability of error localization in the subspace and extended subspace methods at different CEQNRs for a $(10, 5)$ DFT code.
   The curves for the extended case are based on 3 additional syndrome samples, implying that
   the code rate is increased from 0.5 to 0.8.}
  \label{fig:fig2}
    %\vspace{-.05cm}
\end{figure}

\begin{figure}
  \centering
 \includegraphics [scale=0.488] {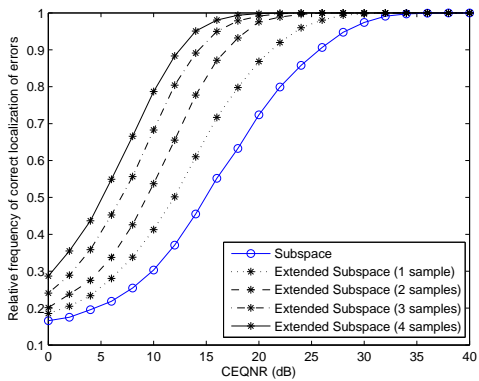}
  %\centering{\epsfig{figure=IC.eps, width=8cm}}
  \caption{Probability of correct localization of
   2 errors for a $(10, 5)$ DFT code using the subspace method and the extended subspace method with
   different number of additional syndrome samples. The the code rate is increased from 0.5 to 0.9 by a step of  0.1.}
  \label{fig:fig3}
    \vspace{-.3cm}
\end{figure}

In Fig.~\ref{fig:fig2}, we compare the frequency of correct localization of errors
for the subspace and extended subspace approaches given a $(10, 5)$ code for different errors.
The gain due to the extended subspace method is remarkable both for one and two errors;
it is more significant for two errors. In fact, as discussed in Section~\ref{sec:SS},
for $\nu = t$ the subspace approach loses its degrees of freedom (DoF) and its performance drops to that of
the coding-theoretic approach. Providing some extra DoF, at the expense of a higher code rate, the extended subspace approach
significantly improves the error localization.
Figure~\ref{fig:fig3} shows how error localization boosts up when the
additional syndrome samples are involved one by one.
This allows doing DSC using DFT codes in a rate-adaptive
manner.


\begin{figure}
  \centering
 \includegraphics [scale=0.492] {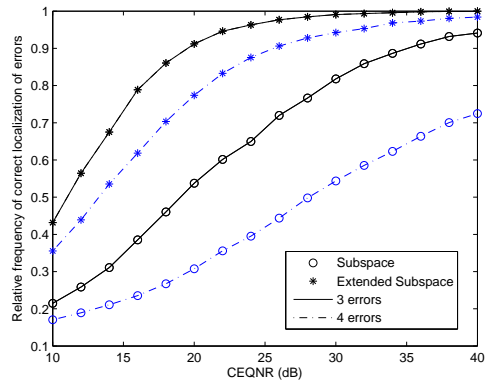}
  %\centering{\epsfig{figure=IC.eps, width=8cm}}
  \caption{Probability of error localization in the subspace and extended subspace methods at different CEQNRs for a $(17,9)$ DFT code.
   The curves for the extended case are based on 4 additional syndrome samples.}
  \label{fig:fig4}
    \vspace{-.1cm}
\end{figure}

The gain caused by the extended subspace method increases for codes
with higher capacity. For instance, simulation results for a $(17, 9)$ DFT code,
presented in Fig.~\ref{fig:fig4}, show a significant gain in any CEQNR
between $10$ to $40 $dB; this is achieved by sending 4 additional syndrome samples.

It is also worth mentioning that numerical results proves the superiority of
using $\tilde R'_m$,
instead of $\tilde R_m$ for finding the number of errors.
Finally, since a better error localization implies a lower reconstruction
error \cite{Vaezi2012WZ}, rate-adapted DFT codes with extended subspace
decoding can be used both to adapt the channel variations and decrease the MSE in DSC.

\section{Conclusion}
\label{sec:con}
We developed two algorithms for rate-adaptation in the DSC system that uses DFT codes for binning.
Rate-adaptation is realized by puncturing the parity samples in the parity-based DSC,
or augmenting the syndrome samples in the syndrome-based DSC. For decoding,
we introduced an extension of subspace error localization algorithm
that substantially improves the error detection and localization,
for a slight increase in the code rate. Interestingly, the gain caused by the extended subspace approach
increases when capacity of the code or the number of errors  go up.
While the algorithm was successfully applied to the syndrome-based DSC in general, we
have been able to exploit it only for the codes with rate $0.5$ in the parity-based system.
The extended decoding algorithm can be applied to DCT and DST codes, as well.
%One possible future research direction would be to investigate applying this algorithm to channel coding as well.

%\section{Appendix}

%\typeout{}
%\bibliography{ThesisEx}
%\bibliographystyle{ieeetr}

\end{document}